\documentclass[12pt]{article}
\usepackage{xspace}
\usepackage{pgf, pgfplots}

\usepackage[margin=1in]{geometry}
\usepackage[utf8]{inputenc}
\usepackage{amsmath}
\usepackage{amsthm}
\usepackage{amssymb}
\usepackage{csquotes}
\usepackage{verbatim}
\usepackage{tikz}
\usepackage{mathtools}
\usepackage{algorithm}
\usepackage{algorithmicx}
\usepackage[noend]{algpseudocode}
\usepackage{graphicx}
\usepackage[export]{adjustbox}
\usepackage{todonotes}
\usepackage[utf8]{inputenc}

\usepackage{thmtools}

\usetikzlibrary{decorations.pathreplacing,intersections}
\usepackage{subfigure}
\usepackage[hidelinks]{hyperref}
\usepackage[nameinlink]{cleveref}

\newcommand{\vercoprob}{\textsc{Vertex Cover}\xspace}
\newcommand{\fvsetprob}{\textsc{Feedback Vertex Set}\xspace}
\newcommand{\dvercoprob}{\textsc{Delayed} \vercoprob}
\newcommand{\dfvsetprob}{\textsc{Delayed} \fvsetprob}
\newcommand{\dvercoprobres}{\dvercoprob \emph{problem with reservations}\xspace}
\newcommand{\dfvsetprobres}{\dfvsetprob \emph{problem with reservations}\xspace}

\newcommand{\fvset}{feedback vertex set\xspace}

\newcommand{\ttsub}{$2$-$3$-subgraph\xspace}

\newcommand{\R}{\mathbf{R}}

\theoremstyle{definition} 
\newtheorem{definition}{Definition}
\theoremstyle{plain} 
\newtheorem{theorem}{Theorem}
\newtheorem{lemma}{Lemma}
\newtheorem{corollary}{Corollary}

\newtheorem{theoremduplicate}{Theorem}
\newtheorem{lemmaduplicate}{Lemma}

\usepackage{refcount}

\begin{document}
  \title{Delaying Decisions and Reservation Costs}

\date{RWTH Aachen University, Germany}
  \author{
   Elisabet Burjons$^1$
   \and
   Fabian Frei$^2$
   \and
   Matthias Gehnen$^3$
   \and
   Henri Lotze$^3$
   \and
   Daniel Mock$^3$
   \and
   Peter Rossmanith$^3$
  }
  \date{$^1$ York University, Canada\\
          {\footnotesize\texttt{burjons@yorku.ca}}\\[5mm]
        $^2$ ETH Z{\"u}rich, Switzerland\\
          {\footnotesize\texttt{fabian.frei@inf.ethz.ch}}\\[5mm]
        $^3$ RWTH Aachen University, Germany\\
          {\footnotesize\texttt{\{gehnen,lotze,mock,rossmani\}@cs.rwth-aachen.de}}}


  \maketitle 

\begin{abstract}
We study the \fvsetprob and the \vercoprob 
problem in a 
natural 
variant of the classical online 
model that allows for \emph{delayed decisions}
and \emph{reservations}. Both problems can be 
characterized by an obstruction set of subgraphs that 
the online graph needs to avoid. In the case of the \vercoprob 
problem, the obstruction set
consists of an edge (i.e., the graph of two adjacent vertices), 
while for the \fvsetprob problem, the obstruction set contains all cycles. 

In the \emph{delayed-decision} model, an algorithm needs to maintain a
valid partial solution after every request, thus allowing it to
postpone decisions until the current partial solution is no longer
valid for the current request.

The \emph{reservation} model grants an online algorithm the new and
additional option to pay a so-called reservation cost for any given
element in order to delay the decision of adding or rejecting it until
the end of the instance.

For the \fvsetprob problem, we first analyze the variant with only
delayed decisions, proving a lower bound of $4$ and an upper bound of
$5$ on the competitive ratio. Then we look at the variant with both
delayed decisions and reservation. We show that given bounds on the competitive
ratio of a problem with delayed decisions impliy lower and upper
bounds for the same problem when adding the option of reservations.
This observation allows us to give a lower bound of
$\min{\{1+3\alpha,4\}}$ and an upper bound of $\min{\{1+5\alpha,5\}}$
for the \fvsetprob problem.  Finally, we show that the online Vertex
Cover problem, when both delayed decisions and reservations are
allowed, is $\min{\{1+2\alpha, 2\}}$-competitive, where $\alpha \in
\R_{\geq 0}$ is the reservation cost per reserved vertex.  
\end{abstract}

\section{Introduction}

In contrast to classical offline problems, where an algorithm is 
given the entire instance it must then solve, 
an online algorithms has no advance knowledge about the instance it needs to 
solve.  
Whenever a new element of the instance is given, some irrevocable decision 
must be taken before the next piece is revealed.

An online algorithm tries to optimize an objective function that is
dependent on the solution set formed by its decisions.  The {\it
strict competitive ratio} of an algorithm, as defined by Sleator and
Tarjan~\cite{SleatorT85}, is the worst-case ratio of the performance of
an algorithm compared to that of an optimal solution computed by an offline 
algorithm for the
given instance, over all instances.  The competitive ratio of an online
\emph{problem} is then the best competitive ratio over all online
algorithms. For a general introduction to online
problems, we refer to the books of Borodin and Ran
El{-}Yaniv~\cite{BorodinE1998} and of Komm~\cite{KommBook2016}.

Not all online problems admit a competitive algorithm (i.e., one whose 
competitive ratio is bounded by a constant) under the classical model. In 
particular, this is the case for the problems \vercoprob and \fvsetprob
discussed in this paper. 

The goal in the general \vercoprob problem is, given a graph $G=(V,E)$, to 
find a 
minimum set of vertices $S\subseteq V$ such that $G[V \setminus S]$ contains no 
edges, 
i.e., the the obstruction set is a path of length 1.

In the classical online version of \vercoprob, the graph is revealed vertex by 
vertex, including all induced edges, and an online
algorithm must immediately and irrevocably decide for each vertex whether to 
add it to the proposed
vertex cover or not. 

The goal of the \fvsetprob problem is, given a graph $G=(V,E)$, to find a 
minimum set of vertices $S\subseteq V$ such that $G[V \setminus S]$ contains no 
cycles.
In this case, the obstruction set contains all cycles.

In both problems, the non-competitiveness is easy to see:  
If the first vertex is added to the solution set, the instance stops and thus leaving a single-vertex instance with an optimal solution size of zero.
On the other hand, not selecting the first vertex will lead to an instance where this vertex becomes a central vertex, 
either of a star at \vercoprob, or of a friendship graph at \fvsetprob.

These adversarial strategies are arguably pathological and unnatural, as  
decisions are enforced that are not based in the properties of the very problem 
to be solved: 
We need to start constructing a vertex cover before any edge
is presented or a feedback vertex set without it being clear if there are even 
any cycles in the instance.
To address this issue in general online Node- and Edge-Deletion
problems, Komm et al.~\cite{Komm2016} introduced the \emph{preemptive}
online model, which was re-introduced by Chen et al.~\cite{Chen2021} as
the \emph{delayed-decision} model. This model allows an online algorithm to 
remain idle
until a ``need to act'' occurs, which in our case means waiting until
a graph from the obstruction set appears in the online graph. The
online algorithm may then choose to delete any vertices in the current online 
graph. The main remaining restriction is that an
online algorithm may not undo any of these deletions.

\begin{definition}
  Let $G$ be an online graph induced by
  its nodes $V(G) = \{v_1,\ldots,v_n\}$, ordered by their occurrence
  in an online instance. The \dvercoprob problem is to select, for every $i$, a 
  subset of vertices $S_i
  \subseteq \{v_1,\ldots,v_i\}$ with $S_1 \subseteq\ldots\subseteq S_n$ such 
  that the induced subgraph $G[\{v_1,\ldots,v_i\}\setminus S_i]$ contains no 
  edge. The goal is to 
  minimize $|S_n|$.
\end{definition}
The definition of the \dfvsetprob problem is identical, except that 
``contains no edge'' is replaced by ``is cycle-free.''

A constant competitive ratio of 2 for the \dvercoprob problem is simple to 
prove and given in the introduction of the paper by Chen et 
al.~\cite{Chen2021}.
The \dfvsetprob problem, in contrast, is more involved. We show that no 
algorithm can admit a competitive ratio better than 4 and adapt results by 
Bar-Yehuda et al.~\cite{Bar-Yehuda98} to give an algorithm that is strictly
5-competitive as an upper bound.

We also consider the model where decisions can be delayed even further
by allowing an algorithm to \emph{reserve} vertices (or edges) of an instance.  
If removing the reserved vertices from the instance would mean that a valid 
solution is maintained, the instance continues. Once an instance has ended, 
the algorithm can freely select the vertices to be
included in the final solution (in addition to the already irrevocably 
chosen ones) among all presented vertices, regardless 
of their reservation 
status.
This reservation is not free: When computing the final competitive ratio, the 
algorithm has to pay a constant $\alpha \in\R_{\geq 0}$ for each reserved 
vertex; these costs are then added to the size of the chosen solution set.
\begin{definition}
  Let $\alpha \in \R_{\ge 0}$ be a constant and $G$ an online graph 
  induced by its nodes $V(G) = \{v_1,\ldots,v_n\}$, ordered by their
  occurrence in an online instance. The \dvercoprobres
  is to select, for every $i$, vertex subsets $S_i, R_i \subseteq
  \{v_1,\ldots,v_i\}$ with $S_1
  \subseteq\ldots\subseteq S_n$ and $R_1 \subseteq\ldots\subseteq R_n$ such 
  that $G[\{v_1,\ldots,v_i\}\setminus (S_i \cup
  R_i)]$ contains no edge. The goal is to minimize the sum $|S_n| + |T| + 
  \alpha|R_n|$, where $T \subseteq V(G)$ is a minimal vertex subset such that 
  $G - (S_n \cup T)$ contains no
  edge.
\end{definition}
Again, the definition for the \dfvsetprobres is identical, except for replacing 
``contains no edge'' with ``is cycle-free.''

For reservation costs of
$\alpha=0$, the problem becomes equivalent to the offline version, whereas for  
$\alpha\ge1$ taking an element directly into the solution 
set becomes
strictly better than reserving it, rendering this reservation option useless.
The results for \dvercoprob and \dfvsetprob, each with reservations, are depicted in Figure~\ref{fig:vcbounds}.

\begin{figure}
\centering
  \begin{tikzpicture}[scale=0.7]
    \begin{axis}[ ymin=1,xmin=0,xmax=1.2,ymax=2.1, xlabel={Reservation cost 
    $\alpha$}, y label style={below=.0ex},ylabel=Competitive Ratio, grid=major]
      \addplot[name path=b1,domain=0:1/2] {1+2*x};
      \addplot[name path=b2, domain=1/2:1.2] {2};
    \end{axis}
    \node at (3.4,-1.4) {\dvercoprob};
  \end{tikzpicture} \hfill
    \begin{tikzpicture}[scale=0.7]
    \begin{axis}[ ymin=1,xmin=0,xmax=1.2,ymax=5.1, xlabel={Reservation cost 
    $\alpha$}, y label style={below=1.3ex}, ylabel=Competitive Ratio, 
    grid=major]
      \addplot[name path=b1,domain=0:4/5] {1+5*x};
      \addplot[name path=b2, domain=4/5:1.2] {5};
      \addplot[name path=b3, dashed, domain=0:1] {1+3*x};
      \addplot[name path=b4, dashed, domain=1:1.2] {4};
    \end{axis}
    \node at (3.4,-1.4) {\dfvsetprob};
  \end{tikzpicture} 
\caption{Upper and lower bounds on the competitive ratio of \dvercoprob (left) 
and \dfvsetprob (right), each with reservations.}
\label{fig:vcbounds}
\end{figure}

The reservation model is still relatively new and has been applied to the simple knapsack problem~\cite{Boeckenhauer2021} and the secretary
problem~\cite{Burjons2021}. We note that the two cited papers 
consider relative reservation costs, while for the two 
problems in the present paper the cost per item are fixed.

The online \vercoprob problem has not received a lot of attention in the past
years.  Demange and Paschos~\cite{Demange2005} analyzed the online \vercoprob
problem with two variations of how the online graph is revealed:  
either vertex by vertex or in clusters, per induced subgraphs of 
the final graph. The
proven competitive ratios are functions on the maximum degree of 
the graph.
Zhang et al.~\cite{Zhang2020} looked at a variant called the Online 
3-Path
\vercoprob problem, where every induced path on three vertices 
needs to be covered.
In this setting, the competitive ratio is again dominated by the maximum degree of the
graph.
Buchbinder and Naor~\cite{Buchbinder09} considered online integral 
and fractional covering problems formulated as linear programs 
where the covering constraints arrive online. As these are a strong 
generalization of the online \vercoprob problem, they achieve 
only logarithmic and not constant competitive ratios.

There has been some work on improving upon the bound of $2$ for 
some special cases of \vercoprob in the model with delayed 
decisions (under different names). For the \vercoprob problem on 
\emph{bipartite} graphs where one side is offline, Wang and Wong 
\cite{Wang15} give an algorithm achieving a competitive ratio of 
$\frac{1}{1 - 1/e} \approx 1.582$.
Using the same techniques they achieve a competitive ratio of 1.901 
for the full online \vercoprob problem for bipartite graphs and 
for the online fractional \vercoprob problem on general graphs.

To the best of our knowledge, the \fvsetprob problem has received 
no attention in the online setting so far, most likely due to the 
fact that there is no competitive online algorithm for this problem 
in the classical setting. The offline \fvsetprob problem, however,  
has been extensively studied, 
especially in the approximation setting.
One notable algorithm is the one in the paper of Bar-Yehuda et
al.~\cite{Bar-Yehuda98}, yielding an approximation ratio of $4-2/n$ 
on an undirected, unweighted graph. We adapt their notation in
Section~\ref{sec:fvs}, and our delayed-decision algorithm with a 
competitive ratio of $5$ is based on their aforementioned 
approximation algorithm. 
The currently best known approximation ratio of 2 by an (offline)  
polynomial-time algorithm 

was given by Becker and Geiger~\cite{Becker1996}.

The paper is organized as follows.
We first look at the \dfvsetprob problem, giving a 
lower bound of 4 and an upper bound of 5 on the competitive ratio.
Then, we discuss how bounds on obstruction set problems 
without reservation imply bounds on the equivalent problems 
with reservations and vice versa, and how this applies to the
\dfvsetprob problem.
Finally, we consider the \dvercoprobres, giving tight bounds dependent on the reservation costs. 

\newpage

\section{Feedback Vertex Set with Delayed Decisions}\label{sec:fvs}
In this section, we consider the \dfvsetprob problem, 
which is concerned with finding the smallest subset of the vertices
of a graph such that their removal yields a cycle-free graph.
We give almost matching bounds on the competitive ratio 
in the delayed decision model.

\begin{theorem}[Lower Bound]\label{thm:fvs-lb}
Given an $\varepsilon
>0$, there is no algorithm for \dfvsetprob achieving a
competitive ratio of $4-\varepsilon$.
\end{theorem}

\begin{figure}[t]
\centering
    \begin{tikzpicture}[every loop/.style={min distance=10mm},
        scale=0.7,yscale=0.7,every node/.style={scale=0.7}]
  \draw (0,0) circle[radius=2];
  \draw (1,{sqrt(3)}) arc (120:-120:2);
  \draw (3,{sqrt(3)}) arc (120:-120:2);
  \draw (5,{sqrt(3)}) arc (120:-120:2);
  \draw (7,{sqrt(3)}) arc (120:-120:2);

  \draw[fill=red] (-2,0) circle(3pt);
  \draw[fill=red] (1,{sqrt(3)}) circle(3pt);
  \draw[fill=red] (3,{-sqrt(3)}) circle(3pt);
  \draw[fill=red] (5,{sqrt(3)}) circle(3pt);
  \draw[fill=red] (7,{sqrt(3)}) circle(3pt);

  \draw [gray,dashed] (1,{-sqrt(3)}) to [bend left=10] (3,{sqrt(3)});
  \draw [gray,dashed] (1,{-sqrt(3)}) to [bend right=10] (3,{sqrt(3)});

  \draw [gray,dashed] (1,{-sqrt(3)}) to [bend right=20] (7,-{sqrt(3)});
  \draw [gray,dashed] (1,{-sqrt(3)}) to [bend right=40] (7,-{sqrt(3)});

  \node[] (loop1) at (3,{sqrt(3)}) {};
  \path (loop1) edge [loop below, blue, dashed] node {} (loop1);

  \draw [gray,dashed] (5,{-sqrt(3)}) to [bend left=10] (7,-{sqrt(3)});
  \draw [gray,dashed] (5,{-sqrt(3)}) to [bend right=10] (7,-{sqrt(3)});

  \draw [gray,dashed] (3,{sqrt(3)}) to [bend left=10] (5,{-sqrt(3)});
  \draw [gray,dashed] (3,{sqrt(3)}) to [bend right=10] (5,{-sqrt(3)});

  \node[] (loop2) at (7,{-sqrt(3)}) {};
  \path (loop2) edge [loop above, blue, dashed] node {} (loop2);

\draw (1,-2.5) node {A};
\draw (3,-2.5) node {B};
\draw (5,-2.5) node {A};
\draw (7,-2.5) node {B};
\draw (9,-2.5) node {A};
\end{tikzpicture}
\caption{Sketch of the lower bound graph, revealed from left to right. The marked vertices are deleted by an algorithm. Then dashed edges (gray) and finally the self-loops (blue) force more deletions. The competitive ratio tends to 4 with increasing instance size.}
\label{fig:fvs-lowerbound}
\end{figure}
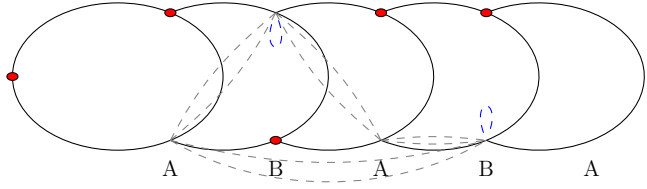

\begin{proof}
The adversarial strategy depicted in 
Figure~\ref{fig:fvs-lowerbound} provides the lower bound.

First, a cycle is presented, which forces any algorithm to
delete a single vertex. The adversary now repeats the following scheme $n$ times:
Identify a pair of vertices of the latest added (half) cycle that 
are still
connected and close a cycle between those two vertices by adding a 
half cycle
between them.

Every time such a half cycle is added, any algorithm has to delete 
a vertex from
it. This is sketched in Figure~\ref{fig:fvs-lowerbound} by the 
black cycles with red dots for exemplary deletions of some 
algorithm. We assume 
w.l.o.g.\ that an algorithm chooses one of the vertices of 
degree 3 between two cycles for deletion, we call these vertices of degree 3 branchpoints. If an algorithm deletes vertices of degree 2 instead, the adversary can force the algorithm to delete even more vertices.

After following this scheme, we split the remaining branchpoints (one for each
new cycle) into two sets $A$ and $B$ alternatingly. We now take each $(a,b)$-pair with $a \in A$ and $b \in B$ and connect the two vertices with two paths, forming a new cycle between each pair of branching vertices. In order to remove all cycles with the lowest amount
of deletions, any algorithm has to delete either all vertices from 
the set $A$ or all vertices from the set $B$. Once such a set is 
chosen, say $A$, the adversary adds ``self-loops'', i.e., new 
cycles each only connected to a vertex of $B$, forcing the deletion 
of all (branchpoint) vertices of $B$.

The optimal solution consists only of the vertices of $B$ and possibly, and
depending on whether $n$ is even and on the choice of $A$ or $B$, 
another
vertex from the first and the last cycle.

Thus, any online algorithm has to delete at least $4n -2$ vertices, while an optimal algorithm only
deletes at most $n + 2$ vertices. The resulting 
competitive ratio is hence worse than $4-\varepsilon$ for large enough values of $n$.
\end{proof}

Next, we present Algorithm~\ref{alg:sub23}, which guarantees a 
competitive ratio of 5. It is a modified version of the 
4-approximative algorithm for \fvsetprob by Bar-Yehuda et 
al.~\cite{Bar-Yehuda98}.
Algorithm~\ref{alg:sub23} maintains a maximal so-called \ttsub $H$ 
in the 
presented graph 
$G$ and selects a feedback vertex set $F$ for $G$ within $H$. It is important to note that $H$ is not necessarily an \emph{induced} subgraph of $G$.
\begin{definition}[\boldmath\ttsub]
	A \emph{\ttsub} of a graph $G$ is a subgraph $H$ of $G$ where 
	every vertex has degree exactly $2$ or $3$ in $H$.
\end{definition}

Given a \ttsub $H$, a vertex is called a \emph{branchpoint of $H$} if it has exactly degree 3 in $H$.
We say a vertex $v$ is a \emph{linkpoint in $H$} if it has exactly degree 2 in $H$ and there is a cycle in $G$ whose only intersection with $H$ is $v$.
A cycle is called \emph{isolated in $H$} if every vertex of this cycle (in $G$) is contained in $H$ and has exactly degree 2 in $H$.

Algorithm~\ref{alg:sub23} adds every branchpoint, linkpoint and one vertex per isolated cycle without linkpoints in $H$ to the solution set $F$.

Note that it is possible that vertices, that were once added to $F$ as linkpoints or due to isolated cycles, can still become branchpoints of degree $3$ in $H$ due to new vertices in $G$. Also later on in the analysis we consider them as branchpoints.

\begin{algorithm}
\caption{Online SubG-2-3}
\begin{algorithmic}
	\State$H \leftarrow \emptyset$, $F \leftarrow \emptyset$
	\For{every new edge in $G$}
		\If{$H$ is not a maximal \ttsub}
		\State Extend $H$ to a maximal \ttsub, converting 
		linkpoints to branchpoints
		\State whenever possible.

    \EndIf
	\State$F \leftarrow F \; \cup$ All new branchpoints
	\State $F\leftarrow F \; \cup$ All new linkpoints
	\State $F\leftarrow F \; \cup$ A set of one vertex per 
	new isolated cycle without linkpoints in $H$
	\EndFor
	\Return $F$
\end{algorithmic}
\label{alg:sub23}
\end{algorithm}

\begin{lemma}[Correctness of Online SubG-2-3]
	Algorithm~\ref{alg:sub23} returns a \fvset for the given input graph $G$.
\end{lemma}
\begin{proof}
By contradiction, assume there is a cycle $C$ in $G$ without a vertex in $F$. 

If $C$ contains no point of $H$, then the complete cycle $C$ can be added to the subgraph $H$ as an isolated cycle, thus $H$ is not maximal, which contradicts the procedure of Algorithm~\ref{alg:sub23}.

Therefore, we can assume that there is at least one vertex of $C$ in $H$. If the cycle contains no branchpoints, either the cycle is an isolated cycle, where all vertices are in $H$ and one vertex is added to $F$, 
or it intersects with $H$ at just a single vertex. 
This vertex is, thus, a linkpoint and part of $F$ as well.
Or, a third option is that $C$ intersects $H$ on two or more points and none of them are branchpoints, in which case, we can extend the \ttsub, which also contradicts the procedure of Algorithm~\ref{alg:sub23}.

Thus, every cycle in $G$ has to contain at least one vertex of $F$, which is a \fvset at the end.
\end{proof}
Proving that Algorithm~\ref{alg:sub23} is $5$-competitive is more tricky.
First, note that an optimal \fvset does not change if we consider a \emph{reduction graph} $G'$ instead of a graph $G$. 
\begin{definition}[Reduction Graph]
	A \emph{reduction graph} $G'$ of a graph $G$ is obtained by 
	deleting vertices of degree $1$ and their incident edges, and
 by deleting vertices of degree $2$, connecting the two neighbors 
 directly (unless the vertex is self-looped).
\end{definition}

The following lemma bounds the size of a reduced graph by its 
maximum degree and the size of any feedback vertex set.
This will be used in the analysis of Algorithm~\ref{alg:sub23}. 
The lemma is due to Voss~\cite{Voss68} and is used in the proof of 4-competitiveness by Bar-Yehuda et al.~\cite{Bar-Yehuda98}.

\begin{lemma}\label{lem:fvssizelim}
	Let $G$ be a graph where no vertex has degree less than $2$. Then for every \fvset $F$, that contains all vertices of degree $2$,
	\[ |V(G)| \leq (\Delta(G)+1)|F|-2 \] holds, where $\Delta(G)$ is the maximum degree of $G$. In particular, if every vertex has a degree of at most $3$, $|V(G)| \leq 4|F|-2$ holds.
\end{lemma}

\begin{theorem}
  Algorithm~\ref{alg:sub23} achieves a strict competitive ratio of $5-2/|V(G)|$ for the online \fvsetprob with delayed decisions.
\end{theorem}
\begin{proof}
	At the end of the instance, after running Algorithm~\ref{alg:sub23}, call the set of branchpoints $B\subseteq F$, the set of linkpoints $L\subseteq F$, and the set of vertices added due to isolated cycles $I$. Again note that vertices can become branchpoints in later steps, even if they were added to $F$ as a linkpoint or for an isolated cycle.
    Let $\mu$ be the size of an optimal \fvset for the graph $G$.
	
	The vertices in $I$ are part of pairwise independent cycles.
    This follows from the fact that every cycle, that was handled as 
    an isolated cycle, was added completely to $H$, thus new isolated cycles cannot contain a vertex of one already added to $H$.
    Therefore, $|I| \leq \mu$, since an optimal solution must contain at least one vertex for each of the pairwise independent cycles.
	
	Moreover, the set of cycles due to to which the vertices of $L$ were added to $F$ are also pairwise independent. This is only true due to the possibility of relabeling linkpoints to branchpoints:
   For a contradiction, first assume that two 
	cycles 
	overlap at a vertex $v$ and and intersect with $H$ at 
	linkpoints $\ell_1$ and $\ell_2$. Then, $H$ could be extended with a 
	path from $\ell_1$ through $v$ to $\ell_2$, which would make $\ell_1$ 
	and $\ell_2$ branchpoints, contradicting the assumption. 
	 Since $\ell_1$ and $\ell_2$ are not branchpoints, both have degree $2$ in $H$ and a new path through the two cycles can be added to $H$, if there are no other points of $H$ on this path. In this case the path would be added to $H$, therefore $\ell_1$ and $\ell_2$ would have become branchpoints. This contradicts the assumption.
  
  If there are other vertices in $H$ on the mentioned path, call the first of those vertices along the path $x$ and assume, w.l.o.g.\ 
	 that it is part of the same cycle as $\ell_1$.
	 If $x$ is connected to either $\ell_1$ or $\ell_2$ via $H$ and the 
	 degree of $x$ in $H$ is at most 2 at any point while it was part of $H$, then we have a 
	 contradiction since $H$ can be extended and one of the linkpoints $\ell_1$ or $\ell_2$ become a 
	 branchpoint. 
 	 
The only case that remains is where $x$ immediately had degree 3 with respect 
to $H$ when it was first added to $H$. In this case $x$ 
is a branchpoint. Note that $\ell_1$ was already deleted as a 
linkpoint at this time, since by definition the cycle would not 
have caused any deleted linkpoints otherwise. 
Since the adversary presents the instance vertex-wise, there must 
be a vertex $s$ such that there was no possibility to add $x$ to 
$H$ before $s$ was presented, but such that $x$ immediately became 
a branchpoint when $s$ was added to $G$. In particular, $x$ cannot have been added immediately to $H$ when it was presented.
Therefore, there must be at least two 
independent paths in $G\setminus H$ from $x$ to $s$. But since the 
algorithm could also add two independent paths from $x$ to $s$  to 
$H$, and one path from $x$ to $\ell_1$, and the algorithm is forced 
to  convert linkpoints to branchpoints whenever possible, it 
adds the path from $x$ to $\ell_1$ first. Note that priority is unambiguous, since there are no paths from $x$ to some other linkpoints in $G\setminus H$: Otherwise $\ell_1$ and the other linkpoint would already be connected within $H$, thus making them branchpoints.
     	 Thus we have $|L| \leq \mu$.
This also shows that there cannot be a branchpoint inside a cycle 
that was used to delete a linkpoint, without also converting the 
linkpoint to a branchpoint.

	It follows that $|L| + |I| \leq 2 \mu$.
	 If $|B| \leq 2 |L|$, then we have $|F| = |I|+|L|+|B| \leq 3 
	 |L| + |I| \leq 4 \mu$, which proves the statement. 
	 
	 In every other case, assume $|B| > 2|L|$.
	 We now consider a reduction graph $H'$ of the graph $H\setminus L$,

	 and delete every component consisting of only a single vertex. 
	 Every vertex in the resulting graph has degree $3$. In 
	 the graph $H$ we can have up to $2|L|$ more branchpoints than 
	 in the resulting graph here.

	 By Lemma~\ref{lem:fvssizelim}, $|B|-2|L|$ is less than $4 \mu (H') -2$, where $\mu(H')$ is the optimal size of a \fvset for the graph $H'$. 
	 
	 The size of an optimal \fvset of the original graph $G$ must 
	 be at least $\mu(H') + |L|$ since every linkpoint in $G$ is 
	 part of a cycle that is not intersecting $H'$. 
	 
	 The inequality chain $|F| = |I|+|L|+|B|  \leq |I|+3|L| + 
	 |B|-2|L| \leq |I| + 4|L| + 4\mu(H')-2 \leq |I| + 4\mu(G)-2 
	 \leq 5 \mu(G)-2$ concludes the proof.

\end{proof}

One could think that the reason Algorithm~\ref{alg:sub23} does not match the competitive ratio of 4 is because the algorithm deletes vertices even in cases where it is not necessary. However, this is not the case. In the appendix we prove that Algorithm~\ref{alg:sub23} cannot be better than $5$-competitive even if vertices in $F$ are only deleted whenever they are part of a completed cycle.

\begin{lemma}\label{lem:delete_in_completed_cycles_only}
The competitive ratio of Algorithm~\ref{alg:sub23} is at least $5-2/|V(G)|$, even if vertices are only deleted whenever necessary.
\end{lemma}

\newpage

\section{Adding Reservations to The Delayed-Decision Model}

We now extend the previous results for the delayed-decision model 
without reservations by presenting two general theorems that translate both upper and lower bounds to the model with reservation.

The delayed-decision model allows us to delay
decisions free of cost as long as a valid solution is maintained. 
Combining delayed decisions with reservations, we have to
distinguish minimization and maximization:
For a minimization problem the default is that the union of selected
vertices and reserved ones constitutes a valid solution at any 
point. 
For a maximization problem, in contrast, it is that the selected 
vertices 
without the
reserved ones always constitute a valid solution. 

Whether the goal is minimizing or maximizing, any 
algorithm for a problem without reservations is also an
algorithm for the problem with reservations,
it just never uses this third option. 
We present a slightly smarter approach for any \emph{minimization} problem such 
as the \dfvsetprob problem. 

\begin{theorem}\label{thm:without_to_with_reservation_upper}
In the delayed-decision model, a $c$-competitive algorithm 
for a minimization problem without reservation yields a 
$\min\{c,1+c\alpha\}$-competitive algorithm for the variant with 
reservation.
\end{theorem}
\begin{proof}
Modify the $c$-competitive algorithm such that it reserves whatever
piece of the input it would usually have immediately selected until
the instance ends -- incurring an additional cost $c\cdot \alpha\cdot
Opt$ -- and then pick an optimal solution for a cost of $Opt$.  This
already provides an upper bound of $1+c\alpha$ on the competitive
ratio.  But running the algorithm without modification still yields an
upper bound of $c$ of course.  
The algorithm can now choose the better of these two options based on
the given $\alpha$, yielding an algorithm that is
$\max\{c,1+c\alpha\}$-competitive.
\end{proof}

\begin{corollary}\label{cor:without_to_with_reservation_upper}
  There is a $\min\{5,1+5\alpha\}$-competitive algorithm
  for the \dfvsetprobres.

\end{corollary}

\begin{theorem}\label{thm:without_to_with_reservation_lower}
In the delayed-decision model, a lower bound of $c$ on the 
competitive ratio for a minimization problem without reservations 
yields a lower bound of 
$\min\{1+(c-1)\alpha,c\}$ on the competitive ratio for the 
problem with reservation.
\end{theorem}
\begin{proof}
The statement is trivial for $\alpha\ge1$; we thus consider now the
case of $\alpha<1$.  Assume that we have an algorithm with
reservations with a competitive ratio better than $1+(c-1)\alpha$.
Even though this algorithm has the option of reserving, it must select
a definitive solution, at the latest when the instance ends. This
definitive solution is of course at least as expensive as the optimal
solution. Achieving a competitive ratio better than $1+(c-1)\alpha$ is
thus possible only if less the algorithm is guaranteed to reserve
fewer than $c-1$ input pieces in total.  But in this case, we can
modify the algorithm with reservation such that it immediately accepts
whatever it would have only reserved otherwise. This increases the
incurred costs by $1-\alpha$ for each formerly reserved input piece,
yielding an algorithm without reservations that achieves a competitive
ratio better than $1+(c-1)\alpha+(c-1)(1-\alpha)=c$.
\end{proof}

\begin{corollary}
 There is no algorithm solving the \dfvsetprobres
 that can achieve a lower bound better than $\min\{1+3\alpha,4\}$.
\end{corollary}

\section{Vertex Cover}

As already mentioned, the \dvercoprob has a competitive ratio of 2 without reservation. 
We now present tight bounds for all reservation-cost values $\alpha$, beginning with the upper bound.

\begin{theorem}\label{thm:vc_reservation_upper}
 There is an algorithm for the \dvercoprobres that achieves a 
 competitive ratio
 of $\min\{1+2\alpha, 2\}$ for any reservation value.
\end{theorem}

These upper bounds, depicted in Figure~\ref{fig:vcbounds}, have 
matching lower bounds.
We start by giving a lower bound for $\alpha\le \frac{1}{2}$.

\begin{theorem}\label{thm:vclb}
 Given an $\varepsilon>0$, there is no algorithm for the \dvercoprobres achieving
 a competitive ratio of $1+2\alpha -\varepsilon$ for any $\alpha\le \frac12$.
\end{theorem}

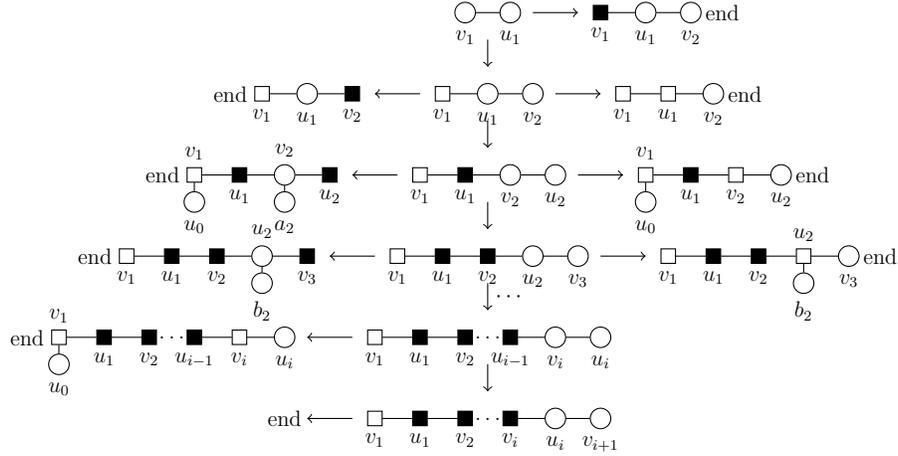
\begin{figure}[h]
    \centering
    \begin{tikzpicture}[node distance=0.7cm,
        cover/.style={draw,rectangle,fill=black},tempco/.style={draw,rectangle},
        vertex/.style={draw,circle},
        scale=0.6,yscale=0.6,every node/.style={scale=0.7}
        ]
        \node[vertex, label=below:$v_1$] (v_1) at (0,0) {};
        \node[vertex, label=below:$u_1$] (u_1) at (1,0) {};
        \draw (v_1) edge (u_1);
        \draw[->] (1.5,0) -- (2.5,0);
        \node[cover,label=below:$v_1$] (v_12) at (3,0) {};
        \node[vertex,label=below:$u_1$] (u_12) at (4,0) {};
        \node[vertex,label=below:$v_2$] (v_22) at (5,0) {};
        \draw (v_12) edge (u_12);
        \draw (v_22) edge (u_12);
        \node at (5.7,0) {end};

        \draw[->] (0.5,-1) -- (0.5,-2);
        \node[tempco,label=below:$v_1$] (v_13) at (-0.5,-3) {};
        \node[vertex,label=below:$u_1$] (u_13) at (0.5,-3) {};
        \node[vertex,label=below:$v_2$] (v_23) at (1.5,-3) {};
        \draw (v_13) edge (u_13);
        \draw (v_23) edge (u_13);
        \draw[->] (2,-3)--(3,-3);
        \node[tempco,label=below:$v_1$] (v_14) at (3.5,-3) {};
        \node[tempco,label=below:$u_1$] (u_14) at (4.5,-3) {};
        \node[vertex,label=below:$v_2$] (v_24) at (5.5,-3) {};
        \draw (v_14) edge (u_14);
        \draw (v_24) edge (u_14);
        \node at (6.2,-3) {end};
        \draw[->] (-1,-3)--(-2,-3);
        \node[tempco,label=below:$v_1$] (v_15) at (-4.5,-3) {};
        \node[vertex,label=below:$u_1$] (u_15) at (-3.5,-3) {};
        \node[cover,label=below:$v_2$] (v_25) at (-2.5,-3) {};
        \draw (v_15) edge (u_15);
        \draw (v_25) edge (u_15);
        \node at (-5.2,-3) {end};

        \draw[->] (0.5,-4)--(0.5,-5);
        \node[tempco,label=below:$v_1$] (v_16) at (-1,-6) {};
        \node[cover,label=below:$u_1$] (u_16) at (0,-6) {};
        \node[vertex,label=below:$v_2$] (v_26) at (1,-6) {};
        \node[vertex,label=below:$u_2$] (u_26) at (2,-6) {};
        \draw (v_16) edge (u_16);
        \draw (v_26) edge (u_16);
        \draw (v_26) edge (u_26);
        \draw[->] (-1.5,-6)--(-2.5,-6);
        \node[tempco,label=above:$v_1$] (v_17) at (-6,-6) {};
        \node[cover,label=below:$u_1$] (u_17) at (-5,-6) {};
        \node[vertex,label=above:$v_2$] (v_27) at (-4,-6) {};
        \node[cover,label=below:$u_2$] (u_27) at (-3,-6) {};
        \node[vertex,label=below:$u_0$] (u_07) at (-6,-7) {};
        \node[vertex,label=below:$a_2$] (a_27) at (-4,-7) {};
        \draw (v_17) edge (u_17);
        \draw (v_27) edge (u_17);
        \draw (v_27) edge (u_27);
        \draw (v_17) edge (u_07);
        \draw (v_27) edge (a_27);
        \node at (-6.7,-6) {end};
        \draw[->] (2.5,-6)--(3.5,-6);
        \node[tempco,label=above:$v_1$] (v_18) at (4,-6) {};
        \node[cover,label=below:$u_1$] (u_18) at (5,-6) {};
        \node[tempco,label=below:$v_2$] (v_28) at (6,-6) {};
        \node[vertex,label=below:$u_2$] (u_28) at (7,-6) {};
        \node[vertex,label=below:$u_0$] (u_08) at (4,-7) {};
        \draw (v_18) edge (u_18);
        \draw (v_28) edge (u_18);
        \draw (v_28) edge (u_28);
        \draw (v_18) edge (u_08);
        \node at (7.7,-6) {end};

        \draw[->] (0.5,-7)--(0.5,-8);
        \node[tempco,label=below:$v_1$] (v_19) at (-1.5,-9) {};
        \node[cover,label=below:$u_1$] (u_19) at (-0.5,-9) {};
        \node[cover,label=below:$v_2$] (v_29) at (0.5,-9) {};
        \node[vertex,label=below:$u_2$] (u_29) at (1.5,-9) {};
        \node[vertex,label=below:$v_3$] (v_39) at (2.5,-9) {};
        \draw (v_19) edge (u_19);
        \draw (v_29) edge (u_19);
        \draw (v_29) edge (u_29);
        \draw (v_39) edge (u_29);
        \draw[->] (3,-9)--(4,-9);
        \node[tempco,label=below:$v_1$] (v_110) at (4.5,-9) {};
        \node[cover,label=below:$u_1$] (u_110) at (5.5,-9) {};
        \node[cover,label=below:$v_2$] (v_210) at (6.5,-9) {};
        \node[tempco,label=above:$u_2$] (u_210) at (7.5,-9) {};
        \node[vertex,label=below:$v_3$] (v_310) at (8.5,-9) {};
        \node[vertex,label=below:$b_2$] (b_210) at (7.5,-10) {};
        \draw (v_110) edge (u_110);
        \draw (v_210) edge (u_110);
        \draw (v_210) edge (u_210);
        \draw (v_310) edge (u_210);
        \draw (b_210) edge (u_210);
        \node at (9.2,-9) {end};
        \draw[->] (-2,-9)--(-3,-9);
        \node[tempco,label=below:$v_1$] (v_111) at (-7.5,-9) {};
        \node[cover,label=below:$u_1$] (u_111) at (-6.5,-9) {};
        \node[cover,label=below:$v_2$] (v_211) at (-5.5,-9) {};
        \node[vertex,label=above:$u_2$] (u_211) at (-4.5,-9) {};
        \node[cover,label=below:$v_3$] (v_311) at (-3.5,-9) {};
        \node[vertex,label=below:$b_2$] (b_211) at (-4.5,-10) {};
        \draw (v_111) edge (u_111);
        \draw (v_211) edge (u_111);
        \draw (v_211) edge (u_211);
        \draw (v_311) edge (u_211);
        \draw (b_211) edge (u_211);
        \node at (-8.2,-9) {end};

        \draw[->] (0.5,-10)--(0.5,-11);
        \node at (1,-10.5) {$\cdots$};
        \node[tempco,label=below:$v_1$] (v_112) at (-2,-12) {};
        \node[cover,label=below:$u_1$] (u_112) at (-1,-12) {};
        \node[cover,label=below:$v_2$] (v_212) at (-0,-12) {};
        \node at (0.5,-12) {$\hdots$};
        \node[cover,label=below:$u_{i-1}$] (u_212) at (1,-12) {};
        \node[vertex,label=below:$v_{i}$] (v_312) at (2,-12) {};
        \node[vertex,label=below:$u_{i}$] (u_312) at (3,-12) {};
        \draw (v_112) edge (u_112);
        \draw (v_212) edge (u_112);
        \draw (v_312) edge (u_212);
        \draw (v_312) edge (u_312);

        \draw[->] (-2.5,-12)--(-3.5,-12);
        \node[vertex,label=below:$u_0$] (u_014) at (-9,-13) {};
        \node[tempco,label=above:$v_1$] (v_114) at (-9,-12) {};
        \node[cover,label=below:$u_1$] (u_114) at (-8,-12) {};
        \node[cover,label=below:$v_2$] (v_214) at (-7,-12) {};
        \node at (-6.5,-12) {$\hdots$};
        \node[cover,label=below:$u_{i-1}$] (u_214) at (-6,-12) {};
        \node[tempco,label=below:$v_{i}$] (v_314) at (-5,-12) {};
        \node[vertex,label=below:$u_{i}$] (u_314) at (-4,-12) {};
        \draw (v_114) edge (u_014);
        \draw (v_114) edge (u_114);
        \draw (v_214) edge (u_114);
        \draw (v_314) edge (u_214);
        \draw (v_314) edge (u_314);
        \node at (-9.7,-12) {end};

        \draw[->] (0.5,-13)--(0.5,-14);
        \node[tempco,label=below:$v_1$] (v_115) at (-2,-15) {};
        \node[cover,label=below:$u_1$] (u_115) at (-1,-15) {};
        \node[cover,label=below:$v_2$] (v_215) at (0,-15) {};
        \node at (0.5,-15) {$\hdots$};
        \node[cover,label=below:$v_{i}$] (v_315) at (1,-15) {};
        \node[vertex,label=below:$u_{i}$] (u_315) at (2,-15) {};
        \node[vertex,label=below:$v_{i+1}$] (v_415) at (3,-15) {};
        \draw (v_115) edge (u_115);
        \draw (v_215) edge (u_115);
        \draw (v_315) edge (u_315);
        \draw (v_415) edge (u_315);
        \draw[->] (-2.5,-15)--(-3.5,-15);
        \node at (-4,-15) {end};
    \end{tikzpicture}
    \caption{Illustration for the proof of Theorem~\ref{thm:vc_reservation_upper}. 
    Adversarial strategy for $\alpha\le \frac12$. Black squares are 
    vertices irrevocably included into the vertex cover, white 
    square vertices are reserved to be temporarily in the vertex 
    cover, and round vertices are not yet chosen.}
    \label{fig:adv-small-alpha}
\end{figure}

\begin{proof}
 We present the following exhaustive set of adversarial instances as depicted in Figure~\ref{fig:adv-small-alpha}, deferred to the appendix.
 First an adversary presents two vertices $u_1$ and $v_1$ connected to each other. 
 Any algorithm for online \vercoprob with reservations is 
 forced to cover this edge either by irrevocably choosing one 
 vertex for the cover or by placing one of the vertices in the 
 temporary cover (i.e., reserving it). In the first case, assume 
 w.l.o.g.\ that $v_1$ is the 
 chosen vertex for the cover. The adversary then presents a vertex 
 $v_2$ connected only to $u_1$ and ends the instance. Such an 
 algorithm would have a competitive ratio of $2$ as it must then 
 cover the edge $(u_1,v_2)$ by placing one of its endpoints in the 
 cover. Choosing vertex $u_1$ alone would have been 
 optimal, however. This is the same lower bound as given for the 
 model without reservations.
 
 If, again w.l.o.g., the vertex $v_1$ is temporarily covered 
 instead, the adversary still presents a vertex $v_2$ connected to 
 $u_1$. Now an algorithm has four options to cover the edge 
 $(u_1,v_2)$: Each of the two vertex $u_1$ or $v_2$ can be either 
 irrevocably chosen or temporarily reserved. If $u_1$ or $v_2$ are 
 temporarily covered, the instance will end here and the 
 reservation costs of the algorithm will be $2\alpha$. Both the 
 algorithm and the optimal solution will end up choosing only 
 vertex $u_1$,
 which implies a final competitive ratio of $1+2\alpha$ in this 
 case. If vertex $v_2$ is chosen, the instance will also end,
 yielding a competitive ratio 
 worse than $2$.
 
 Thus, the only option remaining is to irrevocably cover the vertex 
 $u_1$. In this case, the adversary presents a vertex $u_2$ 
 connected to $v_2$. An algorithm can then irrevocably or 
 temporarily take $v_2$ or $u_2$ respectively. If an algorithm 
 temporarily takes $v_2$ or $u_2$, the adversary will present one 
 more vertex $u_0$ connected to $v_1$ and end the instance. This 
 results in a graph that can be minimally covered by the vertices 
 $v_1$ and $v_2$. The algorithm, however, will have $3$ vertices in 
 the cover and additional reservation costs of $2\alpha$ for the 
 temporarily chosen vertices. Thus it will have a competitive ratio 
 of $\frac32+\alpha$, 
 which is larger than $1+2\alpha$ for the considered values of 
 $\alpha$.
 If an algorithm irrevocably takes $u_2$, the same vertex $u_0$ 
 will be presented and then the instance will end with another 
 auxiliary vertex $a_2$ connected to $v_2$. An optimal vertex cover 
 would take vertices $v_1$ and $v_2$. Any algorithm that 
 has already irrevocably chosen $u_1$ and $u_2$, however, will have 
 to choose 
 two more vertices in order to cover the edges $\{v_1, u_0\}$ and 
 $\{v_2,a_2\}$; thus, its competitive ratio will be worse than $2$.
 
 Again, the only remaining option is to irrevocably choose the 
 vertex $v_2$, after which an adversary presents a vertex $v_3$ 
 connected to $u_2$. An algorithm may choose to irrevocably or 
 temporarily take the vertex $u_2$ or $v_3$. If an algorithm 
 decides to temporarily take any vertex or 
 irrevocably choose $v_3$, then the adversary presents an auxiliary 
 vertex $b_2$ connected to $u_2$ and ends the request sequence. An 
 optimal vertex cover in this case has size two, containing only 
 the vertices $u_1$ and $u_2$. In the best case, however, such an 
 algorithm has a vertex cover of size $3$ and two temporary covers, 
 thus its competitive ratio will be at best $\frac32+\alpha$, which 
 again is worse than $1+2\alpha$, as already observed.
 
 In general, after irrevocably choosing $u_1,\hdots, 
 u_{k-1}$ and $v_2,\hdots, v_{k-1}$, and temporarily choosing 
 $v_1$, the adversary presents the vertex $u_k$ connected to $v_k$. 
 If an algorithm chooses to reserve any of the endpoints or 
 irrevocably selects $u_i$, then the adversary presents the vertex 
 $u_0$ and ends the request sequence. In this case
 an optimal vertex cover only contains the vertices $v_i$ for every 
 $i=1,\hdots k$, thus it has size $k$. The algorithm, however, will 
 have to take $v_1$ and $v_k$ in addition to the previously 
 irrevocably taken vertices, thus obtaining a vertex cover of 
 size $2k-1$ at best together with two temporarily taken vertices.  
 Thus the 
 competitive ratio is 
 $\frac{2k-1+2\alpha}{k}=2-\frac{1-2\alpha}{k}\ge 1+2\alpha\;,$
 where the inequality holds for every $k\ge1$.
 
 In the other case, after irrevocably choosing vertices 
 $u_1,\hdots,u_{k-1}$ and $v_2,\hdots, v_k$, the adversary presents 
 the vertex $v_{k+1}$ connected to $u_k$. If an algorithm chooses 
 to reserve one of the two endpoints or irrevocably chooses 
 $v_{k+1}$, then the adversary stops the request sequence. 
 An optimal vertex cover of such a graph consists of the vertices 
 $u_i$ for every $i=1,\hdots,k$ and it has size $k$. The 
 algorithm will have to choose $u_i$ in order to obtain a vertex 
 cover at all, obtaining a vertex cover of size $2k-1$ at best 
 together with at least one reservation, thus achieving a 
 competitive ratio of 
 $\frac{2k-1+\alpha}{k}\ge 1+2\alpha - \varepsilon$
 for any $k \geq \frac{1}{\varepsilon}$.
\end{proof}

For larger values of $\alpha$ the same adversarial strategy holds, but it gives us the following lower bound.

\begin{theorem}\label{thm:vclb2}

  For $\alpha > 1/2$, no algorithm for the \dvercoprobres
  is better than 2-competitive.
\end{theorem}

\begin{proof}
  The lower bound of Theorem~\ref{thm:vclb} for $\alpha = 1/2$ is $2 - \varepsilon$. For larger values of $\alpha$ 
  the same adversarial strategy will give us a lower bound of $2$. This is because, at all points during the analysis, either the value of the competitive ratio for each strategy was at least $2$, or it had a positive correlation with the value of $\alpha$, meaning that for larger values of $\alpha$ any algorithm following that strategy obtains strictly worse competitive ratios.
\end{proof}

\section{Conclusion}
We have shown that some problems that are non-competitive in the classical model become competitive in modified, but natural
variations of the classical online model. Some questions remain open, such as the best competitive ratio for the \dfvsetprob 
problem, which
we believe to be 4.

It may be worthwhile to investigate which results can be 
found for restricted graph classes.
For example, it is easy to see that the online version of 
\dfvsetprob is 2-competitive on graphs with maximum degree 
three.

In addition we also introduced the reservation model on graphs,
 providing an upper and a lower bound for general graph problems.
 It would be interesting try to find matching bounds, also on specific graph problems.

\bibliography{delayed-node-deletion}

\begin{thebibliography}{10}

\bibitem{Bar-Yehuda98}
Reuven Bar{-}Yehuda, Dan Geiger, Joseph Naor, and Ron~M. Roth.
\newblock Approximation algorithms for the feedback vertex set problem with
  applications to constraint satisfaction and bayesian inference.
\newblock {\em {SIAM} J. Comput.}, 27(4):942--959, 1998.

\bibitem{Becker1996}
Ann Becker and Dan Geiger.
\newblock Optimization of pearl's method of conditioning and greedy-like
  approximation algorithms for the vertex feedback set problem.
\newblock {\em Artif. Intell.}, 83(1):167--188, 1996.

\bibitem{Boeckenhauer2021}
Hans{-}Joachim B{\"{o}}ckenhauer, Elisabet Burjons, Juraj Hromkovi\v{c}, Henri
  Lotze, and Peter Rossmanith.
\newblock Online simple knapsack with reservation costs.
\newblock In {\em {STACS} 2021}, volume 187 of {\em LIPIcs}, pages 16:1--16:18,
  2021.

\bibitem{BorodinE1998}
Allan Borodin and Ran El{-}Yaniv.
\newblock {\em Online computation and competitive analysis}.
\newblock Cambridge University Press, 1998.

\bibitem{Buchbinder09}
Niv Buchbinder and Joseph Naor.
\newblock Online primal-dual algorithms for covering and packing.
\newblock {\em Math. Oper. Res.}, 34(2):270--286, 2009.

\bibitem{Burjons2021}
Elisabet Burjons, Matthias Gehnen, Henri Lotze, Daniel Mock, and Peter
  Rossmanith.
\newblock The secretary problem with reservation costs.
\newblock In {\em {COCOON} 2021}, volume 13025 of {\em LNCS}, pages 553--564,
  2021.

\bibitem{Chen2021}
Li{-}Hsuan Chen, Ling{-}Ju Hung, Henri Lotze, and Peter Rossmanith.
\newblock Online node- and edge-deletion problems with advice.
\newblock {\em Algorithmica}, 83(9):2719--2753, 2021.

\bibitem{Demange2005}
Marc Demange and Vangelis~Th. Paschos.
\newblock On-line vertex-covering.
\newblock {\em Theoretical Computer Science}, 332(1):83--108, 2005.

\bibitem{KommBook2016}
Dennis Komm.
\newblock {\em An Introduction to Online Computation -- Determinism,
  Randomization, Advice}.
\newblock Texts in Theoretical Computer Science. Springer, 2016.

\bibitem{Komm2016}
Dennis Komm, Rastislav Kr{\'a}lovi\v{c}, Richard Kr{\'a}lovi\v{c}, and
  Christian Kudahl.
\newblock {Advice Complexity of the Online Induced Subgraph Problem}.
\newblock In {\em MFCS 2016}, volume~58 of {\em LIPIcs}, pages 59:1--59:13,
  2016.

\bibitem{SleatorT85}
Daniel~Dominic Sleator and Robert~Endre Tarjan.
\newblock Amortized efficiency of list update and paging rules.
\newblock {\em Commun. {ACM}}, 28(2):202--208, 1985.

\bibitem{Voss68}
Heinz-Jürgen Voss.
\newblock Some properties of graphs containing $k$ independent circuits.
\newblock {\em Theory of Graphs. Proceedings of Colloquium Tihany}, pages
  321--334, 1968.

\bibitem{Wang15}
Yajun Wang and Sam~Chiu{-}wai Wong.
\newblock Two-sided online bipartite matching and vertex cover: Beating the
  greedy algorithm.
\newblock In {\em ICALP 2015}, volume 9134 of {\em LNCS}, pages 1070--1081,
  2015.

\bibitem{Zhang2020}
Yubai Zhang, Zhao Zhang, Yishuo Shi, and Xianyue Li.
\newblock Algorithm for online 3-path vertex cover.
\newblock {\em Theory Comput. Syst.}, 64(2):327--338, 2020.

\end{thebibliography}
\newpage

\section*{Appendix}
\subsection*{Deferred Proofs}
For convenience, we restate all statements before proving them.

\setcounterref{lemmaduplicate}{lem:delete_in_completed_cycles_only}
\addtocounter{lemmaduplicate}{-1}
\begin{lemmaduplicate}
The competitive ratio of Algorithm~\ref{alg:sub23} is at least $5-2/|V(G)|$, even if vertices are only deleted whenever necessary.
\end{lemmaduplicate}
\begin{proof}
The following adversarial strategy, illustrated in Figure~\ref{fig:sub23-lowerbound}, provides the desired lower bound.
  
First, the adversary presents $n$ independent cycles $C_1,...,C_n$. 
Algorithm~\ref{alg:sub23} deletes at least one vertex per cycle.

Second, the adversary connects each pair of neighboring cycles 
$C_i$ and $C_{i+1}$ for $i \leq n-1$ with two independent paths.
This adds $4n-4$ branchpoints. 
Everything that is presented is part of the \ttsub, marked in blue 
in Figure~\ref{fig:sub23-lowerbound}.

Now a vertex $c$ (the top vertex in 
Figure~\ref{fig:sub23-lowerbound}) is added and connected to every 
branchpoint 
with two edges.
This new vertex will not be part of the \ttsub. 
Therefore, even a modified Algorithm~\ref{alg:sub23} that only deletes branchpoints whenever necessary has to delete every branchpoint.
Here, Algorithm~\ref{alg:sub23} deletes $n+4n-4$ vertices, whereas 
an optimal solution consists of only one vertex per independent 
cycle and the vertex $c$.
\begin{figure}[h]
 \centering
      \begin{tikzpicture}[scale=0.9]
      \draw[blue] (0,0) circle[radius=1];
      \draw[blue] (3,0) circle[radius=1];
      \draw[blue] (6,0) circle[radius=1];
      \draw[blue] (9,0) circle[radius=1];

      \draw[blue] (0.5,{sqrt(3)/2}) --  (2.5,{sqrt(3)/2});
      \draw[blue] (3.5,{sqrt(3)/2}) --  (5.5,{sqrt(3)/2});   
      \draw[blue] (6.5,{sqrt(3)/2}) --  (8.5,{sqrt(3)/2});   
      \draw[blue] (0.5,-{sqrt(3)/2}) --  (2.5,-{sqrt(3)/2});   
      \draw[blue] (3.5,-{sqrt(3)/2}) --  (5.5,-{sqrt(3)/2});   
      \draw[blue] (6.5,-{sqrt(3)/2}) --  (8.5,-{sqrt(3)/2});    
   
      \draw [red,dashed] (4.5,2) to [bend left=10] (2.5,{sqrt(3)/2});
      \draw [red,dashed] (4.5,2) to [bend left=10] (0.5,{sqrt(3)/2});
      \draw [red,dashed] (4.5,2) to [bend left=10] (3.5,{sqrt(3)/2});
      \draw [red,dashed] (4.5,2) to [bend left=10] (5.5,{sqrt(3)/2});
      \draw [red,dashed] (4.5,2) to [bend left=10] (6.5,{sqrt(3)/2});
      \draw [red,dashed] (4.5,2) to [bend left=10] (8.5,{sqrt(3)/2});

      \draw [red,dashed] (4.5,2) to [bend left=10] (2.5,-{sqrt(3)/2});
      \draw [red,dashed] (4.5,2) to [bend left=10] (0.5,-{sqrt(3)/2});
      \draw [red,dashed] (4.5,2) to [bend left=10] (3.5,-{sqrt(3)/2});
      \draw [red,dashed] (4.5,2) to [bend left=10] (5.5,-{sqrt(3)/2});
      \draw [red,dashed] (4.5,2) to [bend left=10] (6.5,-{sqrt(3)/2});
      \draw [red,dashed] (4.5,2) to [bend left=10] (8.5,-{sqrt(3)/2});

      \draw [red,dashed] (4.5,2) to [bend right=5] (2.5,{sqrt(3)/2});
      \draw [red,dashed] (4.5,2) to [bend right=5] (0.5,{sqrt(3)/2});
      \draw [red,dashed] (4.5,2) to [bend right=5] (3.5,{sqrt(3)/2});
      \draw [red,dashed] (4.5,2) to [bend right=5] (5.5,{sqrt(3)/2});
      \draw [red,dashed] (4.5,2) to [bend right=5] (6.5,{sqrt(3)/2});
      \draw [red,dashed] (4.5,2) to [bend right=5] (8.5,{sqrt(3)/2});

      \draw [red,dashed] (4.5,2) to [bend right=5] (2.5,-{sqrt(3)/2});
      \draw [red,dashed] (4.5,2) to [bend right=5] (0.5,-{sqrt(3)/2});
      \draw [red,dashed] (4.5,2) to [bend right=5] (3.5,-{sqrt(3)/2});
      \draw [red,dashed] (4.5,2) to [bend right=5] (5.5,-{sqrt(3)/2});
      \draw [red,dashed] (4.5,2) to [bend right=5] (6.5,-{sqrt(3)/2});
      \draw [red,dashed] (4.5,2) to [bend right=5] (8.5,-{sqrt(3)/2});

      \draw[fill=red] (4.5,2) circle(2pt);
      \draw[fill=blue] (0.5,{sqrt(3)/2}) circle(2pt);
      \draw[fill=blue] (3.5,{sqrt(3)/2}) circle(2pt);   
      \draw[fill=blue] (6.5,{sqrt(3)/2}) circle(2pt);   
      \draw[fill=blue] (0.5,-{sqrt(3)/2}) circle(2pt);   
      \draw[fill=blue] (6.5,-{sqrt(3)/2}) circle(2pt);   
      \draw[fill=blue] (3.5,-{sqrt(3)/2}) circle(2pt);   
      \draw[fill=blue] (2.5,{sqrt(3)/2}) circle(2pt);   
      \draw[fill=blue] (5.5,{sqrt(3)/2}) circle(2pt);   
      \draw[fill=blue] (8.5,{sqrt(3)/2}) circle(2pt);   
      \draw[fill=blue] (2.5,-{sqrt(3)/2}) circle(2pt);   
      \draw[fill=blue] (5.5,-{sqrt(3)/2}) circle(2pt);   
      \draw[fill=blue] (8.5,-{sqrt(3)/2}) circle(2pt);

      \draw[fill=blue] (0,-1) circle(2pt);   
      \draw[fill=blue] (3,-1) circle(2pt);
      \draw[fill=blue] (6,-1) circle(2pt);
      \draw[fill=blue] (9,-1) circle(2pt);
      
    \end{tikzpicture}
    \caption{The \ttsub is given in blue, marked vertices will be
    deleted by Algorithm~\ref{alg:sub23}. The top vertex cannot be added to the
    \ttsub, so it will not be deleted.}
    \label{fig:sub23-lowerbound} 
  \end{figure}
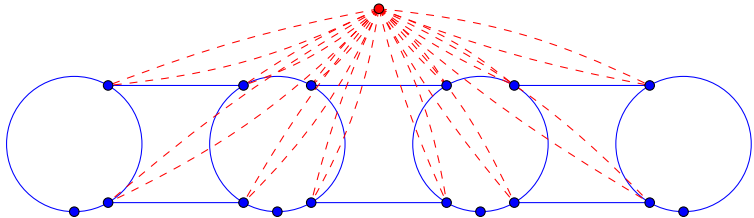
\end{proof}

\setcounterref{theoremduplicate}{thm:vc_reservation_upper}
\addtocounter{theoremduplicate}{-1}
\begin{theoremduplicate}\label{thm:vc_reservation_upper}
  There is an algorithm for the \dvercoprobres that achieves a 
  competitive ratio
  of $\min\{1+2\alpha, 2\}$ for any reservation value.
 \end{theoremduplicate}
 \begin{proof}
 We present an algorithm that achieves a competitive ratio of 
 $1+2\alpha$ and -- together with the algorithm by Chen et 
 al.~\cite{Chen2021}, which does not reserve vertices -- we achieve 
 a competitive ratio of $\min\{1+2\alpha, 2\}$.
 
 Our algorithm is itself an adaptation of the classical 2-approximation for \vercoprob. Given a new vertex, the algorithm considers every edge and whenever an edge is uncovered the algorithm temporarily covers both endpoints by reserving the two vertices. 
 Given a graph~$G$ with a minimal vertex cover
 of size~$k$, this algorithm incurs reservation costs of
 $\alpha\cdot 2k$, as the algorithm selects at most as 
 many vertices as $2\cdot Opt$, where $Opt$ is the size of the 
 optimal vertex cover.
 However, because the final decision on the vertex cover is 
 completely left to the very last step and no vertex is permanently 
 chosen during the running of the algorithm, once the whole instance 
 is presented the algorithm can choose a minimal vertex cover for 
 $G$ as the final solution. Thus, its competitive ratio is 
 $\frac{k+\alpha\cdot 2k}{k}=1+2\alpha$.
 For $\alpha > 1/2$, the online algorithm without reservations by 
 Chen et al.~\cite{Chen2021} has a competitive ratio of~$2$, beating the ratio of~$1+2\alpha$.
 \end{proof}

\subsection*{Deferred Illustration}

\end{document}